\documentclass[aps,pra,twocolumn,superscriptaddress,showpacs,floatfix,nofootinbib,llncs,longbibliography]{revtex4-1}
\usepackage{graphicx}
\usepackage{amssymb}
\usepackage{amsmath}
\usepackage{color}
\usepackage{psfrag}
\usepackage{epsfig}
\usepackage{bbm}
\usepackage{bm}
\usepackage{hyperref}
\usepackage[normalem]{ulem}
\usepackage{amsthm}
\usepackage{epstopdf}
\usepackage{verbatim}
\usepackage{subfigure}
\usepackage[export]{adjustbox}
\usepackage{bbold}

\usepackage{array}
\usepackage{mathtools}

\newcommand{\ket}[1]{| #1 \rangle}

\newcolumntype{M}[1]{>{\centering\arraybackslash}m{#1}}

\begin{document}
\title{Boosting entanglement growth of many-body localization by superpositions of disorder
}
\author{Jhen-Dong Lin}
\email{jhendonglin@gmail.com}
\affiliation{Department of Physics, National Cheng Kung University, 701 Tainan, Taiwan}
\affiliation{Center for Quantum Frontiers of Research \& Technology, NCKU, 70101 Tainan, Taiwan}

\author{Yueh-Nan Chen}
\email{yuehnan@mail.ncku.edu.tw}
\affiliation{Department of Physics, National Cheng Kung University, 701 Tainan, Taiwan}
\affiliation{Center for Quantum Frontiers of Research \& Technology, NCKU, 70101 Tainan, Taiwan}

\date{\today}

\begin{abstract}
Many-body localization (MBL) can occur when strong disorders prevent an interacting system from thermalization. To study the dynamics of such systems, it is typically necessary to perform an ensemble average over many different disorder configurations. Previous works have utilized an algorithm in which different disorder profiles are mapped into a quantum ancilla. By preparing the ancilla in a quantum superposition state, quantum parallelism can be harnessed to obtain the ensemble average in a single computation run. In this work, we modify this algorithm by performing a measurement on the ancilla. This enables the determination of conditional dynamics not only by the ensemble average but also by the quantum interference effect. Using a phenomenological analysis based on local integrals of motion, we demonstrate that this protocol can lead to an enhancement of the dephasing effect and a boost in the entanglement growth for systems in the deep MBL phase. We also present numerical simulations of the random XXZ model where this enhancement is also present in a smaller disorder strength, beyond the deep MBL regime.

\end{abstract}

\maketitle
\section{Introduction}
Thermalization is ubiquitous in non-integrable many-body systems. However, in a seminal work~\cite{AndersonPhysRev}, Anderson demonstrated that strong disorder can restore the integrability and prevent the thermalization of non-interacting systems, a phenomenon termed Anderson localization. Recently, both experimental and theoretical investigations have shown that strong disorders can also localize interacting systems, which is known as many-body localization (MBL)~\cite{nandkishore2015many, Abanin2019rmp, alet2018many}. One of the milestones in this field is the proposal of a phenomenological model called local integrals of motions (LIOMs)~\cite{Serbyn2013prl, Huse2014prb, ros2015integrals, Wahl2017prx, Mierzejewski2018prb, Thomson2018prb}, which characterizes MBL systems with extensive conserved quantities. From the perspective of LIOMs, the slow logarithmic entanglement growth~\cite{Bardarson2012prl, Serbyn2013prl2}--the hallmark of MBL--can be explained by the dephasing mechanism together with the exponential decay law for interactions among LIOMs. This insight has led to experimental proposals such as spin-echo type experiments~\cite{Serbyn2014prl, Serby2017prb, Chiaro2022prr} to identify MBL systems.

To extract meaningful information from such random disordered systems, an ensemble average over many disorder configurations is usually necessary, which could be a resource-intensive task. In Ref.~\cite{Paredes2005prl}, the authors proposed an algorithm that utilizes quantum parallelism to address this issue. The main idea is to encode different disorder profiles into a quantum ancilla. The ancilla can be prepared in a quantum superposition state, allowing the MBL dynamics for different profiles to run in parallel. By tracing out the ancilla at the end, the ensemble average results can be obtained in a single computation or experiment run~\cite{Alvarez2008prl, Andraschko2014prl, papic2015annphys}. 

Recently, the idea of using quantum ancilla to evolve systems in parallel has also been adopted in various disciplines, including quantum communication~\cite{chiribella2019quantum, Loizeau2020pra, abbott2020quantum,
kristjansson2020njp, rubino2021prr}, quantum thermodynamics~\cite{rubino2021communphys, nie2022arxiv, chan2022pra}, quantum metrology~\cite{Chapeaupra2021, lee2023prr, siltanen2022arxiv}, open quantum systems~\cite{ban2021pla, ban2020qip, Siltanen2021pra,lin2022prr}, and relativistic quantum theory~\cite{foo2021prd, foo2020prd, Henderson2020prl}. However, instead of directly discarding the ancilla after an evolution, these works considered performing a measurement on the ancilla. In this case, the conditional state of the system may depend not only on the ensemble average but also on the quantum interference effect among different quantum evolutions, leading to nontrivial results.

Motivated by these works, we modify the algorithm proposed in Ref.~\cite{Paredes2005prl} by performing an additional measurement, as mentioned earlier. Our aim is to investigate the potential impacts of this modified algorithm on MBL systems. We first provide analytical and numerical analysis based on LIOM representation. Our findings show that the quantum interference for different evolutions (disorder profiles) can generally enhance the dephasing effect and boost the entanglement growth for systems in the deep MBL phase. Moreover, we consider the random XXZ model~\cite{xxz2008prb, pal2010prb, luitz2015prb}, the well-investigated model in the context of MBL, with parameters ranging from small to strong disorder strengths. We also consider the case that can manifest Anderson localization. The numerical results demonstrate a significant enhancement of entanglement growth and saturation for both the many-body localization and Anderson localization scenarios. Furthermore, as we increase the number of superposed disorder profiles, the system's memory can be retained, indicating that the mechanism of dephasing enhancement is also valid beyond the deep MBL phase.

The rest of this paper is organized as follows. In sec.~\ref{sec:liom}, we formulate the algorithm and present a phenomenological analysis based on the LIOM representation. In sec.~\ref{sec:XXZ}, we present numerical results for the random XXZ model that can demonstrate either many-body or Anderson localization. Finally, we draw our conclusion in sec.~\ref{sec:conclusion}.

\section{Phenomenological analysis \label{sec:liom}}
\begin{figure*}
    \includegraphics[width=1\linewidth, height=0.4\linewidth]{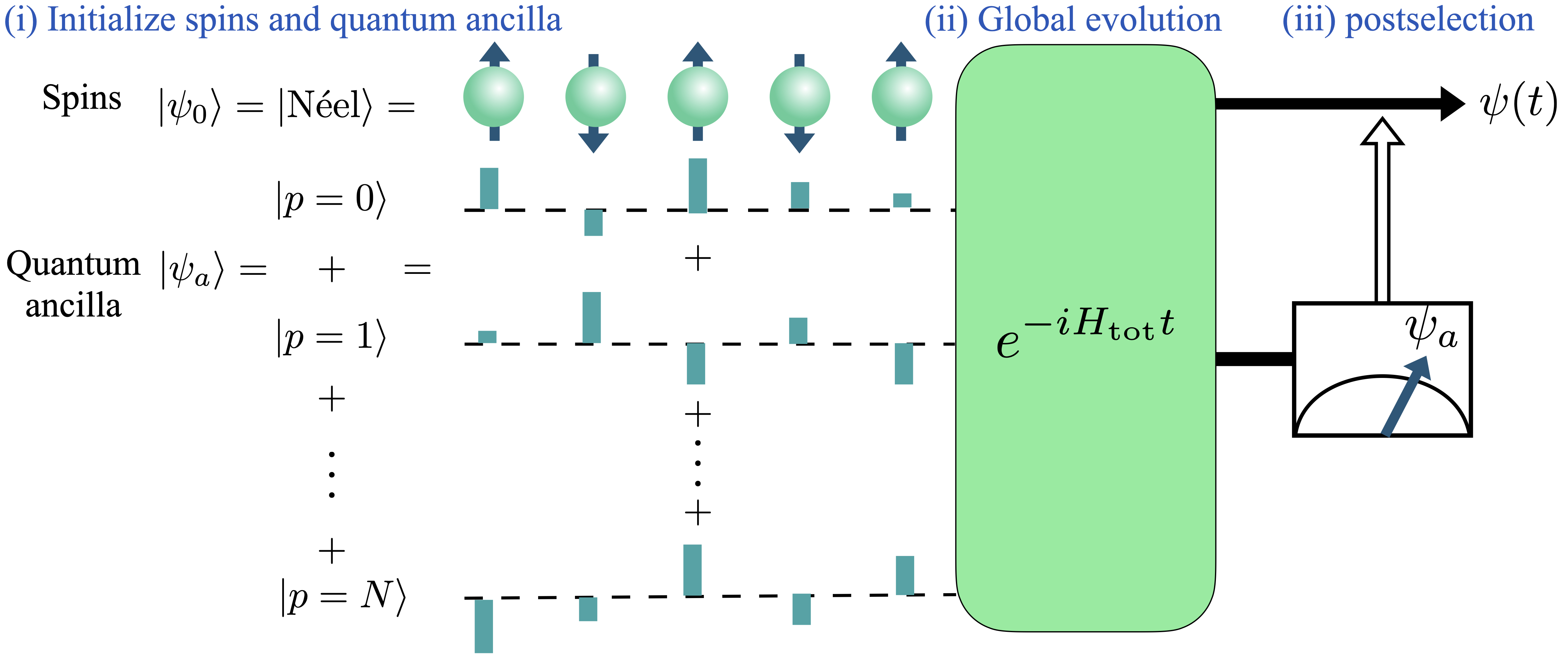}
    \caption{Schematic illustration of the algorithm for spin chain subject to a ‘‘quantum superposition" of random onsite potential (presented by the histograms), where the profiles are encoded in a quantum ancilla. In step (i), we prepare the ancilla in a quantum superposition state $\ket{\psi_a}=\sum_p\ket{p_a}/\sqrt{N}$; that is, the superposition of profiles of onsite potentials. In step (ii), the total system is evolved according to the Hamiltonian defined in Eq.~\eqref{eq:H_tot} to achieve quantum parallel computation. In step (iii), we postselect the spin system associated with the ancilla state $\ket{\psi_a}$, and the conditional state of the spin $\ket{\psi(t)}$ can be obtained.}
    \label{fig:illus}
\end{figure*}

In this section, we consider the well-known phenomenological model for MBL. Specifically, for a spin chain in the MBL phase, there exist extensively conserved quantities that can be obtained by quasi-local unitary transformation $U$ such that
\begin{align}
    \tau_{z,i} &= U^\dagger \sigma_{z,i} U\nonumber\\
    &= \sigma_{z,i} + \sum_{j,k}\sum_{\alpha,\beta=x,y,z}C_{\alpha,\beta}(j,k)\sigma_{\alpha,j}\sigma_{\beta,k}+\cdots. \label{l-bit}
\end{align}
Here, $\{\sigma_{\alpha,i}\}_{\alpha=x,y,z}$ are the usual Pauli operators on site $i$, and $\tau_{z,i}$ is the conserved quantity associated with $\sigma_{z,i}$. The coefficient $C_{\alpha,\beta}(j,k)$ decays exponentially with the distance between the spins $j$ and $k$. Conventionally, $\tau_{z,i}$ and $\sigma_{z,i}$ are called the localized bit (l-bit) and the physical bit (p-bit), respectively. In other words, l-bits can be regarded as p-bits associated with quasi-local dressings. Because these l-bits are conserved, the system's Hamiltonian can be written in general as 
\begin{equation}
    H_{\text{MBL}} = \sum_i h_i \tau_{z,i} + \sum_{i<j}J_{i,j}~\tau_{z,i}\tau_{z,j}+\cdots. \label{H_MBL}
\end{equation}
The interactions between these l-bits are described by an exponential decay law, i.e., $J_{i,j} = \tilde{J}_{i,j}\exp(-|i-j|/\xi )$, where $\xi$ denotes a dimensionless characteristic length.
Here, we consider that the pre-factor $\tilde{J}_{i,j}$ is uniformly drawn form the interval $[-\mathcal{J},\mathcal{J}]$. 

This model captures several dynamical features of a generic MBL system. For instance, the characteristic trait of slow entanglement generation can be described by the dephasing interactions among these l-bits. Specifically, due to the exponential decay of the interaction between two distant l-bits, it takes an exponentially long time to build up entanglement, which results in a logarithmic entanglement growth~\cite{Serbyn2014prl, Serby2017prb, Chiaro2022prr}. As an example, we consider an initial product state:

\begin{equation}
    |\psi_0\rangle = \frac{1}{\sqrt{2}}\left(|\uparrow\rangle + |\downarrow\rangle\right)\otimes \frac{1}{\sqrt{2}}\left(|\uparrow\rangle +|\downarrow\rangle\right)\otimes \cdots , \label{eq:superposition state}
\end{equation}
where $\tau_{z}|\uparrow\rangle= |\uparrow\rangle$ and $\tau_{z}|\downarrow\rangle= -|\downarrow\rangle$. We focus on the reduced dynamics of the first l-bit (i.e. $j=1$). Additionally, we consider a simplified variant of the model Hamiltonian where we only keep the two-body interactions in Eq.~\eqref{H_MBL}:
\begin{align}
    &\tilde{H}_{\text{MBL}} = \sum_{i<j}J_{i,j}\tau_{z,i}\tau_{z,j} \nonumber \\
    &= |\uparrow_1\rangle\langle \uparrow_1|\sum_{j>1} J_{1,j}\tau_{z,j} - |\downarrow_1\rangle\langle \downarrow_1|\sum_{j>1} J_{1,j}\tau_{z,j} + H'\nonumber \\
    &\text{with~}H'=\sum_{1<i<j}J_{i,j}\tau_{z,i}\tau_{z,j}.
\end{align}
In this case, the reduced dynamics of the first l-bit can be expressed as
\begin{align}
    &\rho(t) =\mathrm{tr}_{j>1}\left[\exp\left(-i\tilde{H}_{\text{MBL}}t\right)|\psi_0\rangle \langle \psi_0|\exp\left(i\tilde{H}_{\text{MBL}}t\right) \right] \nonumber\\
    &= \frac{1}{2}\left[|\uparrow\rangle \langle\uparrow| + |\downarrow\rangle \langle \downarrow| + \phi(t) |\uparrow \rangle \langle \downarrow| + \phi(t)^*|\downarrow\rangle \langle \uparrow| \right], 
\end{align}
where the dephasing factor is given by
\begin{equation}  
  \phi(t)=\prod_{j>1}\cos(2J_{1,j}t). \label{eq:dephasing factor}
\end{equation} The dephasing factor directly reflects the entanglement dynamics. For instance, one can quantify the entanglement between the first and other l-bits by linear entanglement entropy, i.e., $S_\text{L}=\left(1 - |\phi(t)|^2\right)/2$.

The dynamics can be divided into two regimes according to the product form in Eq.~\eqref{eq:dephasing factor}: (1) The short-time regime where $\mathcal{J}t\leq \exp(1/\xi)$, and the entanglement generation is governed by the interactions nearby the first l-bit (i.e., $|1-j|<\xi$), resulting in a power-law increase in the entanglement entropy. (2) When $\mathcal{J}t\gg \exp(1/\xi)$, the dynamics is governed by the long-range interactions (i.e., $|1-j|>\xi$), where the effect of the exponential decay factor in the interactions becomes significant and results in the logarithmic entanglement growth. Note that $\phi(t)$ is independent of the evolution governed by $H'$, wherein the interactions do not involve the first l-bit. However, as shown in the following, this is not the case when we consider quantum superpositions of disorders by using the quantum ancilla, which leads to faster entanglement growth.

Let's consider our algorithm, bearing in mind that the explicit expressions of $\tau_{z,i}$ and the diagonalized Hamiltonian described by Eq.~\eqref{H_MBL} depend on the specific parameters of the system, such as a particular disorder configuration. However, in a deep MBL regime, e.g., strong disorder limit, the dressing terms in Eq.~\eqref{l-bit} becomes negligible, and the l-bits and p-bits roughly coincide. Along this reasoning, we expect that different disorder profiles will give rise to different realizations of the pre-factors $\{\tilde{J}_{i,j}\}$ of the diagonalized Hamiltonian, while the explicit expressions of l-bits remain unchanged.

\begin{figure}
\includegraphics[width=1\linewidth]{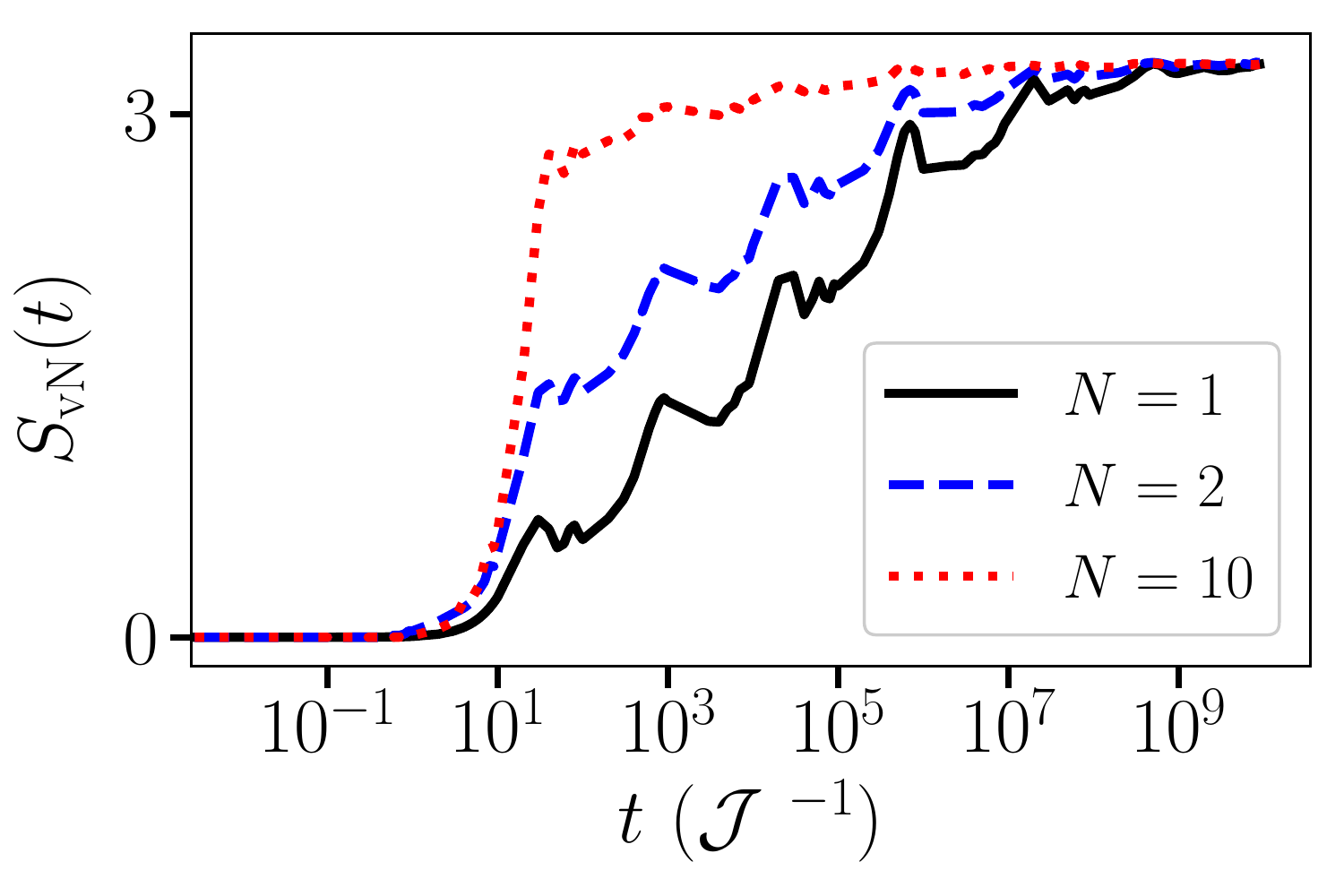}
\caption{Time evolutions of half-chain entanglement of an eight l-bits system with 1, 2, and 10 superposed disorder configurations. Here $\mathcal{J}=1$ and $\xi=0.3$. The results are averaged over 1000 disorder realizations.
\label{liom1}}
\end{figure}

Suppose we encode $N$ different disorder profiles into a quantum ancilla, which can then determine $N$ different evolutions. The ancilla-system total Hamiltonian is given by
\begin{align}
    H_{\text{tot}} = \sum_{p=1}^N|p_a\rangle \langle p_a|\otimes \tilde{H}_{\text{MBL}}(\{J_{i,j}^{(p)}\}).\label{eq:H_tot}
\end{align}
Here, $\tilde{H}_{\text{MBL}}(\{J_{i,j}^{(p)}\})$ denotes the Hamiltonian with a specified realization of the interactions $\{J_{i,j}^{(p)}\}$, and $\{|p_a\rangle\}$ forms an orthonormal basis with $\langle p_a| p'_{a}\rangle =\delta_{p,p'}$. In Fig.~\ref{fig:illus}, we illustrate the algorithm for a spin chain subject to onsite random potential, which can be divided into three steps:
\begin{itemize}
    \item[(i)]  Initialization: We initialize the spin system in state $|\psi_0\rangle$. In this section, we consider the state in Eq.~\eqref{eq:superposition state}. In sec.~\ref{sec:XXZ}, we will consider the \text{N\'{e}el} state, which will be formally introduced later. The quantum ancilla is prepared in a superposition state $|\psi_a\rangle = \sum_p|p_a\rangle /\sqrt{N}$ to achieve quantum superposition of disorder configurations.
    \item[(ii)] Global evolution: We let the ancilla--spin system evolve. The time-dependent state can then be written as 
\begin{equation}
|\psi_{\text{tot}}(t)\rangle=\frac{1}{\sqrt{N}}\sum_p |p_a\rangle\otimes \exp\left[-i \tilde{H}_{\text{MBL}}\left(\{J_{i,j}^{(p)}\}\right)t\right]|\psi_0\rangle.
\end{equation}
    \item[(iii)] Postselection: We postselect the state $|\psi_a\rangle$ on the ancilla before discarding it. After the postselection, the conditional state of the system reads
\begin{align}
    &|\psi(t)\rangle = \frac{|\tilde{\psi}(t)\rangle}{|| \tilde{\psi}(t)||} \nonumber \\
    &\text{with~}|\tilde{\psi}(t)\rangle = \frac{1}{N}\sum_p \exp\left[-i \tilde{H}_{\text{MBL}}\left(\{J_{i,j}^{(p)}\}\right)t\right]|\psi_0\rangle. 
\end{align}
\end{itemize}

The effective dephasing factor of the first l-bit can be described by
\begin{align}
    &\phi_{\text{eff}}(t) \propto \frac{1}{N^2}\sum_{p,q}\langle \psi_0'|\hat{F}_{p,q}(t)\hat{G}_{p,q}(t) |\psi_0'\rangle\nonumber
\end{align}
with
\begin{align}
    &\hat{F}_{p,q}(t) = \exp\left(-i\left[H^{'(p)}-H^{'(q)}\right]t\right), \nonumber\\
    &\hat{G}_{p,q}(t) = \exp\left[-i\sum_{j>1}\left(J_{1,j}^{(p)}+J_{1,j}^{(q)}\right)\tau_{z,j}t\right],\nonumber\\
    &|\psi_0'\rangle = \bigotimes_{j>1}\frac{|\uparrow_j\rangle +|\downarrow_j\rangle}{\sqrt{2}}.
\end{align}
Note that the effective dephasing factor now includes the term $H'$ through the factor $\hat{F}_{p,q}(t)$ with $p\neq q$, which originates from the quantum interference effect~\cite{lin2022prr} between the profiles $p$ and $q$. This factor introduces additional contribution of interactions, i.e., $J_{i,j}$ with and $i,j>1$, into the dephasing factor. Following the intuition of the product form in Eq.~\eqref{eq:dephasing factor}, one can expect that the dephasing effect and the entanglement growth will become stronger. To confirm this intuition, in Fig.~\ref{liom1}, we present numerical simulations of entanglement dynamics for an eight l-bits system with $\mathcal{J}=1$ and $\xi = 0.3$. The entanglement is quantified by the half-chain von-Neumann entropy $S_{\text{vN}} = \mathrm{tr}\rho \log_2 \rho $, where $\rho$ denotes the half-chain reduced density matrix. One can clearly observe that entanglement growth is enhanced as the number of superposed evolutions $N$ increases. Therefore, the results indicate that \textit{the quantum interference effect for dynamics with different disorder profiles can enhance the entanglement growth for a generic system in the deep MBL regime.}

\section{XXZ chain with random external field \label{sec:XXZ}}
\begin{figure}
\includegraphics[width=1\linewidth]{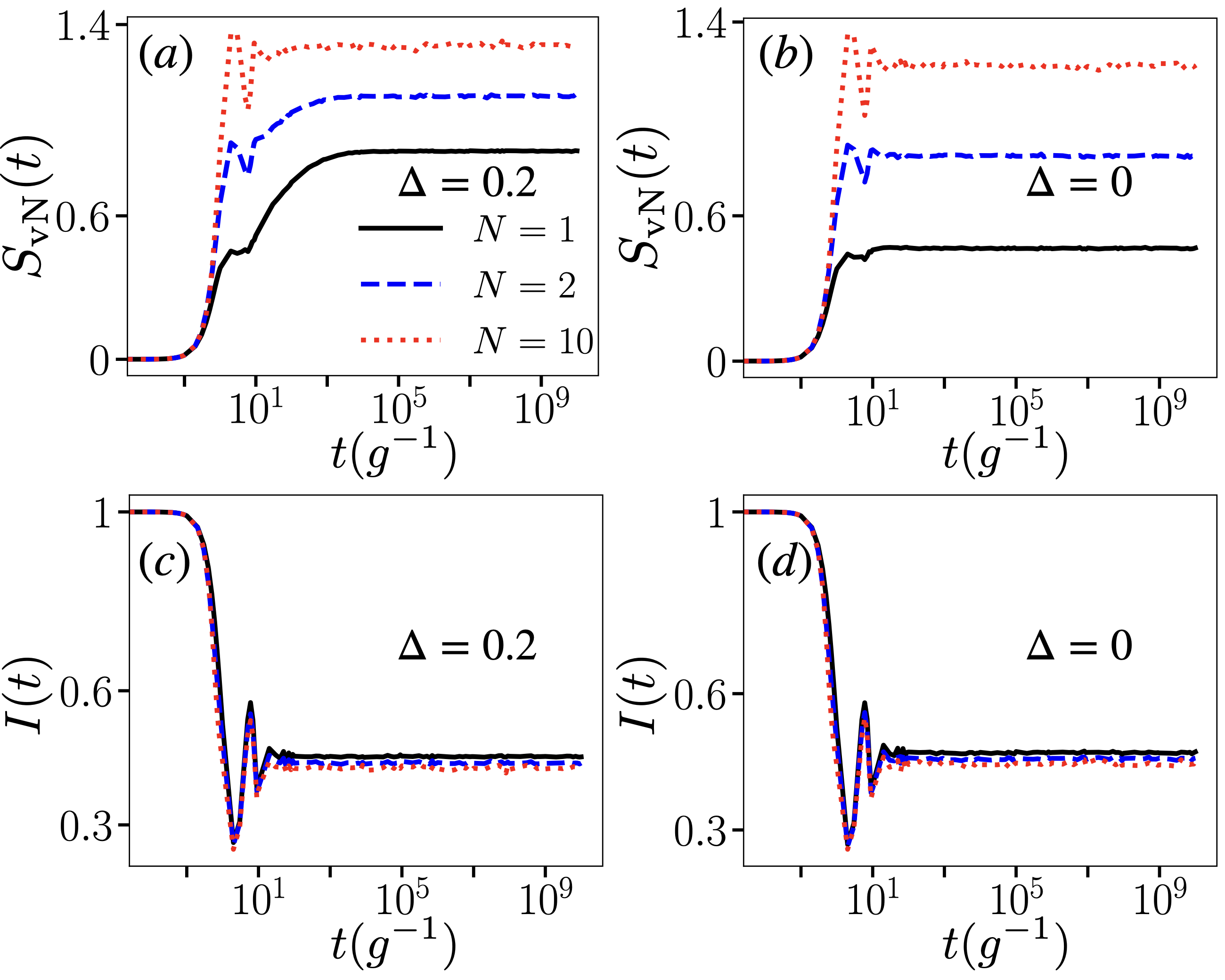}
\caption{Time evolutions of entanglement entropy (a,b) and Imbalance (c,d) for different numbers of profiles $N$ in quantum superposition. The disorder strength is $W=3$, and we consider $\Delta=0.2$ and $\Delta=0$, which can demonstrate many-body localization and Anderson localization, respectively. The results are obtained by averaging over 10000 disorder realizations.}
\label{fig:dynamics}
\end{figure}

\begin{figure}
\includegraphics[width=1\linewidth]{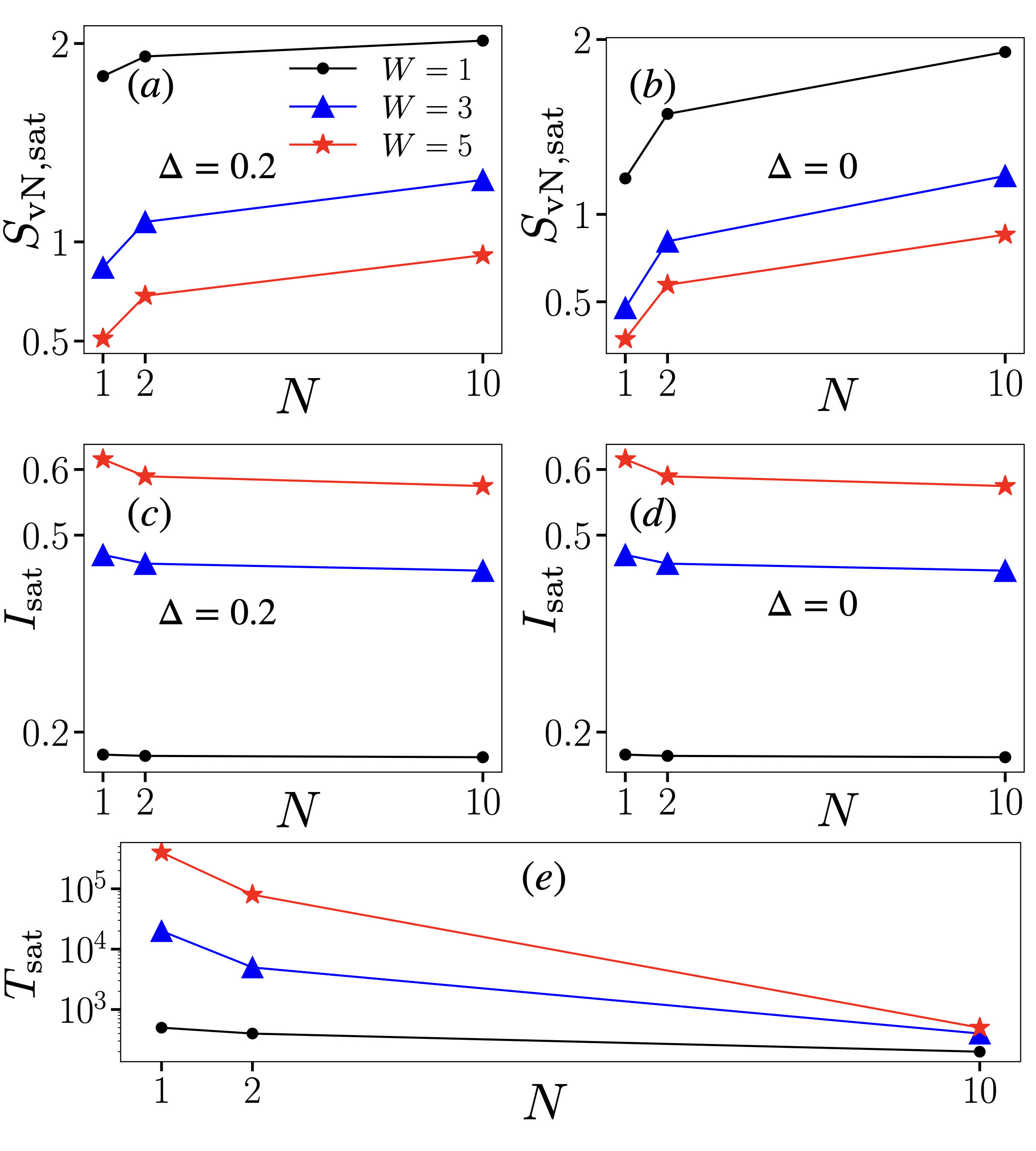}
\caption{(a)--(d) The saturation values of entanglement entropy and imbalance for different disorder strengths and different numbers of superposed profiles. The many-body localization scenario ($\Delta = 0.2$) and Anderson localization scenario ($\Delta = 0$) are considered. (e) The saturation time (for the case $\Delta=0.2$) of the entanglement entropy for different disorder strengths and different numbers of superposed profiles. All of the results are obtained by averaging over 10000 disorder realizations.}
\label{fig:saturation}
\end{figure}

In this section, we consider XXZ Heisenberg chain with random external field~\cite{xxz2008prb, pal2010prb, luitz2015prb}, where the Hamiltonian reads
\begin{align}
    H_{\text{XXZ}} &= \sum_{i}g\left(S_{x,i}S_{x,i+1}+S_{y,i}S_{y,i+1}\right) +\Delta S_{z,i}S_{z,i+1}
    \nonumber\\
    &~~~+\sum_ih_i S_{z,i}.
\end{align}
Here, $\{S_{x,i}=\sigma_{x,i}/2, S_{y,i}=\sigma_{y,i}/2, S_{z,i}=\sigma_{z,i}/2 \}$ denote spin-$1/2$ operators on site $i$. Also, $g$ and $\Delta$ describe the hopping and interaction energies between each spin and their nearest neighbors. The $h_i$ represents the strength of the external field on site $i$ and is uniformly drawn from $[-W,W]$ with $W$ being the disorder strength. When the disorder strength is sufficiently strong, the system can display localization behavior. Note that for the case $\Delta=0$, the system can exhibit Anderson localization, whereas for the case $\Delta\neq 0$, the system can demonstrate many-body localization. In addition to the entanglement dynamics, we also consider the imbalance, which is an experimental relevant quantifier of the system's memory~\cite{schreiber2015science, smith2016natphys, choi2016science, xu2018prl, Doggen2010prb,
chanda2020prr}, as an indicator for the localization behavior. We initialize the system in the N\'{e}el state, i.e,
\begin{equation}
    |\psi_{\text{N\'{e}el}}\rangle = |1,-1,1,-1,\cdots\rangle, 
\end{equation} 
with $\sigma_{z}|\pm 1\rangle =\pm|\pm1\rangle$. The imbalance is defined as 
\begin{equation}
    I(t) = \sum_{i=1} (-1)^{i+1}\langle \psi(t)|\sigma_{z,i}|\psi(t)\rangle.
\end{equation}
Since the imbalance quantifies the amount of the initial memory of the system at the time $t$, one expects that the $I(t)$ decays to zero in an ergodic (thermalized), implying that the initial memory is completely lost. Therefore, the presence of a non-zero value of imbalance in the long time limit suggests that the system is in a localized phase.

In Fig.~\ref{fig:dynamics}, we present the dynamics of the half-chain entanglement $S_{\text{vN}}$ and the imbalance $I(t)$ for an eight-spin chain with different numbers of superposed disorder profiles $N$. Throughout the following discussions, we set the hopping energy $g=1$. We consider both the cases $\Delta = 0.2$ and $\Delta = 0$ and set $W = 3$ so that many-body localization and Anderson localization can be observed. As shown in Fig.~\ref{fig:dynamics} (a), for the case of $\Delta = 0.2$, one can observe a slow logarithmic growth in $S_{\text{vN}}$ followed by an initial fast growth, which is the hallmark of many-body localization. In contrast, for the case of $\Delta =0$ [Fig.~\ref{fig:dynamics} (b)], i.e., Anderson localization, the slow logarithmic growth is absent. One can also observe that increasing the number $N$ can boost the initial entanglement growth and enhance the saturated entanglement for both cases. Furthermore, for the case, $\Delta =0.2$, the duration for the logarithmic growth becomes shorter when increasing the number $N$. However, the enhancement of entanglement does not imply a strong thermalization of the system, as shown in Fig.~\ref{fig:dynamics} (c) and (d), where we find that the dynamics and the saturation value of the imbalance $I(t)$ are relatively insusceptible to the number $N$, suggesting that the system can stay in a localized phase. In other words, the initial memory can be preserved, while generating more entanglement among the spins. Therefore, one can conclude that increasing the number $N$ results in a stronger dephasing effect, as suggested in the previous section.  

In Fig.~\ref{fig:saturation} (a)--(d), we present the saturation values of $S_{\text{vN}}$ and $I$ with different disorder strengths $W=\{1,3,5\}$. The saturation values are defined by 
\begin{align}
    Q_{\text{sat}} = \frac{1}{t_f-t_i}\int_{t_i}^{t_f} dt ~Q(t),
\end{align}
with $Q\in \{S_{\text{vN}},I\}$. Also, we set $t_i/g=10^9$ and $t_f/g = 10^{10}$. In Fig.~\ref{fig:saturation} (e), we consider the saturation time of $S_{\text{vN}}$ (for the case $\Delta = 0.2$), which is defined by
\begin{equation}
    T_{\text{sat}} = \min t ~~\text{s.t.~~} |S_{\text{vN}}(t)-S_{\text{vN,sat}}|<\epsilon.
\end{equation}
Here, we set the tolerance $\epsilon= 10^{-3}$. One can observe that when $W$ decreases, $I_{\text{sat}}$ ($S_{\text{vN,sat}}$) decreases (increases), indicating that the system becomes more thermalized as the disorder strength decreases. Further, for a fixed $W$, one can observe that $S_{\text{vN,sat}}$ enhances when $N$ increases. Particularly, we can observe that the saturation time of entanglement $T_{\text{sat}}$ significantly reduces for a larger disorder strength. For instance, $T_{\text{sat}}$ becomes almost three-order smaller when $N$ increases from $1$ to $10$ for the case $W=5$. Finally, we can find that increasing $N$ does not significantly reduce $I_{\text{sat}}$ even for a smaller disorder strength. It means that the mechanism, enhancing the dephasing effect, also manifests beyond the deep MBL regime for this model.

%\sectio{Experimental proposal \label{sec:exp}}

\section{Conclusions\label{sec:conclusion}}
We consider an algorithm, which can superpose dynamics with different disorder profiles, and investigate its impact on MBL systems. Through a phenomenological analysis, we demonstrate that the interference effect can result in an enhancement of entanglement generation for generic systems in the deep MBL regime. Moreover, we present numerical simulations on the random XXZ model and consider two scenarios that can manifest many-body localization and Anderson localization. For both cases, the long-time saturated entanglement is enhanced. Notably, the superposition of disorders can significantly boost the entanglement growth for the MBL scenario. 

This work raises several open questions. For instance, since quantum superposed dynamics can manipulate the entanglement properties of MBL systems, it is natural to ask about its implications for practical applications, e.g., quantum many-body battery~\cite{rossini2019prb} and quantum computation~\cite{wang2022nc}, etc. Also, from a more general perspective, the proposed algorithm can be regarded as a many-body engineering protocol. Therefore, it would be intriguing to investigate its applications to other exotic many-body effects, such as quantum information scrambling~\cite{landsman2019nc} and quantum time crystal~\cite{zhang2017nature}. 

Regarding the possible experimental implementations, we note that there are existing proposals based on cold atoms~\cite{horstmann2007pra,
krutitsky2008pra,
horstmann2010prl,
PhysRevA.105.013324} and trapped ions~\cite{bermudez2010localization}, enabling one to map disorders into an auxiliary quantum degree of freedom.

\section*{Acknowledgement}
This work is supported by the National Center for Theoretical Sciences and National Science and Technology Council, Taiwan, Grants No. MOST 111-2123-M-006-001.
%\bibliography{ref}


\end{document}